\documentstyle[12pt]{article}
\newcommand{\be}{\begin{equation}}
\newcommand{\bea}{\begin{eqnarray}}
\newcommand{\ee}{\end{equation}}
\newcommand{\eea}{\end{eqnarray}}

\input FEYNMAN

\begin{document}
\renewcommand{\thefootnote}{\fnsymbol{footnote}}
\begin{flushright}
IISc-CTS-13/96\\
TIFR/TH/96-40\\
hep-ph/9607385\\
\end{flushright}
\vskip 30pt
\begin{center}
{\Large \bf 
Transverse structure function in the  
factorisation method} \\
\vspace{5mm}
{\bf Prakash Mathews\footnote{prakash@cts.iisc.ernet.in; Address after August 
96: 
{\it Theory Group, Tata Institute of Fundamental Research, 
Homi Bhabha Road, Bombay-400 005, India.}
}}\\
\vspace{4pt}
{\it Centre for Theoretical Studies, Indian Institute of Science,
Bangalore-560012, India.}\\
\vspace{8pt}
{\bf V. Ravindran\footnote{ravi@prl.ernet.in}}\\
\vspace{4pt}
{\it Theory Group, Physical Research Laboratory, Navrangpura,
Ahmedabad-380 005, India.}\\
\vspace{8pt}
and \\
\vspace{4pt}
{\bf K. Sridhar\footnote{sridhar@theory.tifr.res.in}}\\
\vspace{4pt}
{\it Theory Group, Tata Institute of Fundamental Research, \\ 
Homi Bhabha Road, Bombay-400 005, India.}

\vspace{20pt}
{\bf ABSTRACT}
\end{center}

Deep Inelastic scattering experiments using transversely polarised targets 
yield information on the structure function $g_2$.  By means of a free-field 
analysis, we study the operator structure of $g_2$ and demonstrate the need 
for retaining the twist three mass terms in order to maintain
current-conservation.  
We show that the structure function $g_T=g_1+g_2$ has a much simpler
operator structure as compared to $g_2$, in spite of the fact
that, like $g_2$, $g_T$ has a twist-three component.
We demonstrate factorisation of the hadronic 
tensor into hard and soft parts for the case of $g_T$. 
We show that the first moment of the gluonic contribution 
to $g_T$ vanishes, and discuss possible physical applications.

\setcounter{footnote}{0}
\renewcommand{\thefootnote}{\arabic{footnote}}

\vfill
\clearpage
\pagestyle{plain}

	The recent transversely polarised deep inelastic scattering (DIS) 
experiments \cite{g2e} have opened up new avenues to explore the 
polarised structure of the proton.  With more data, better accuracy and 
new proposed experiments \cite{Jaf-T}, it would soon be possible to 
extract the twist-three contribution for the first time.  Though extraction 
of the twist-three contributions is in general quite difficult\cite{Mul,Jaf-R}, 
in these experiments it is possible to kinematically eliminate the 
leading twist contribution \cite{Jaf-CNP}.

	The polarised structure of the proton is characterised by two 
structure functions $g_1(x,Q^2)$ and $g_2(x,Q^2)$ which can be 
measured in a polarised lepton-proton DIS experiment $\ell (k) P(p)$ $ 
\rightarrow \ell (k^\prime) X(p_x)$.  The spin dependent part of the 
proton tensor is parametrised as
\begin{eqnarray}
\widetilde W_{\mu \nu} (x,Q^2) &=&  \frac{i}{p \cdot q} \epsilon_
{\mu \nu\lambda \sigma} ~q^\lambda \left \{ s^\sigma \left ( g_1 (x,Q^2) 
+ g_2 (x,Q^2) \right ) - \frac{q \cdot s} {p \cdot q} ~p^\sigma g_2 (x,Q^2) 
\right \}~,
\label{WMUNU}
\end{eqnarray}
where $s_\mu$ is the spin vector of the proton and is normalised as $s^2=-M^2$
with $s \cdot p =0$, $M$ being the target mass. The spin-dependent 
cross-section \cite{Jaf-CNP} is given by
\begin{eqnarray}
\frac{d \Delta \sigma (\alpha)} {dx dy d \phi} &=& \frac{e^4}{4 \pi^2 Q^2}
\left \{ \cos \alpha \left \{ \left [ 1-\frac{y}{2} - \frac{y^2}{4} 
(\kappa - 1) \right] g_1 (x,Q^2) - \frac{y}{2} (\kappa-1) g_2 (x,Q^2) 
\right \} \right . \nonumber \\
&-& \left . \sin \alpha \cos \phi \sqrt{(\kappa -1) \left(1-y-\frac{y^2}{4} 
(\kappa-1) \right)} \left [\frac{y}{2} ~g_1 (x,Q^2) + g_2 (x,Q^2) \right ] 
\right \}~,
\label{CS}
\end{eqnarray}
where $y=p \cdot q/p \cdot k$, $\kappa=1+ 4 x^2 M^2/Q^2$, $\phi$ is the 
azimuthal angle and $\alpha$ is angle between the spin 
vector $s$ and the incoming lepton momentum $k$.  
In a longitudinally polarised experiment ($\alpha=0$), the dominant 
contribution comes from the structure function $g_1 (x,Q^2)$ while $g_2
(x,Q^2)$ is suppressed by a factor $M^2/Q^2$, thus enabling the extraction 
of $g_1 (x,Q^2)$. The longitudinally polarised DIS
process has been studied quite extensively \cite{Jaf-R,AEL} and there is a 
considerable amount of data on $g_1 (x,Q^2)$ \cite{g1}.  
In contrast, the extraction of $g_2 (x,Q^2)$ requires transversely 
polarised proton $(\alpha = 90)$ and further this cross-section is 
suppressed by a factor $M/ \sqrt{Q^2}$ relative to the longitudinal case.  
Note that at the cross-section level the transverse asymmetry measures the 
twist-three contribution while the longitudinal asymmetry measures the 
twist-two contribution.  Hence the extraction of $g_2 (x,Q^2)$ is much more 
complicated as compared to $g_1 (x,Q^2)$. Recently, experimental information 
on $g_2(x,Q^2)$ has become available \cite{g2e}, but the data have large 
errors and do not provide a definite answer to the question of the validity 
of the sum-rules associated with $g_2 (x,Q^2)$, like the Burkhardt-Cottingham
(BC) sum-rule \cite{BC}, the Wandzura-Wilczek (WW) sum-rule \cite{WW}, or 
the recently proposed Efremov-Leader-Teryaev (ELT) sum-rule \cite{ELT}.
contribution etc.

	In transversely polarised DIS experiments, the asymmetry that is 
measured is the virtual photon absorption asymmetry 
\begin{eqnarray}
A_2 (x,Q^2) &=& \frac{\sqrt{Q^2}}{\nu} ~ \frac{g_1 (x,Q^2) + g_2 (x,Q^2)}
 {F_1 (x,Q^2)}~,
\label{A2}
\end{eqnarray}
where $F_1 (x,Q^2)$ is the spin-averaged structure function. 
We see from eq.~\ref{A2} that the asymmetry is proportional not to
$g_2$ alone, but to $g_T (x,Q^2) = g_1 (x,Q^2) + g_2 (x,Q^2)$. 
In this letter, we show that the quantity $g_T$ admits of a
much simpler description than does $g_2$. We suggest that this 
may help in going some way towards a fuller understanding of the
transverse spin structure of the nucleon. We begin by discussing
the free field theory analysis \cite{JaJi} in order to elucidate the
operator structure of the structure functions and demonstrate
the importance of the mass term in maintaining gauge 
invariance of the hadronic tensor. We then discuss the first moment 
of $g_T(x,Q^2)$ and its relation to the spin content of the proton, 
and study the gluonic contribution to the first moment, using the
Factorisation Method (FM) \cite{CSS}.

The hadronic tensor $W_{\mu \nu}(p,q,s)$ has the form 
\be
W_{\mu \nu}(p,q,s)={1 \over 4 \pi} \int d^4 \xi~ e^{iq.\xi} 
~ \langle p s|~ [J_\mu(\xi),~J_\nu(0)]~ | p s \rangle_c ~,
\ee
Retaining the dominant contribution in the light-cone
limit, $\xi^2 \rightarrow 0$, identified as the most singular part 
of the time-ordered product of these currents on the light-cone,
we find 
\bea
\widetilde W_{\mu \nu}(p,q,s) &=& {i \over 4 \pi^2} 
\epsilon_{\mu \nu \lambda
\rho} \int d^4 \xi ~ e^{iq.\xi} ~ \xi^\lambda ~ \delta^{(1)}(\xi^2) ~ 
\epsilon(\xi_0) ~ \langle p s \vert :\! {\cal O}^\rho_A(\xi,0)\!: 
\vert p s \rangle_c ~,
\label{WMuNulc}
\eea
where
\be
{\cal O}^\rho_A (\xi,0) =
\bar \psi(\xi) \gamma^\rho \gamma_5 \psi(0)+ 
\bar \psi(0) \gamma^\rho \gamma_5 \psi(\xi)~,
\ee
To arrive at the above result we used 
\bea
iS(\xi,0)&=&-\langle 0\vert T (\psi(\xi) \bar \psi(0) )\vert 0\rangle~,\\
&=& -{i \over 2 \pi^2} {\not \! \xi \over (\xi^2 -i \epsilon)^2} + O(m)~,
\label{prop1}
\eea
where order $m$ terms are neglected.  The importance of
these terms will be shown later.  To find the dominant contribution 
coming from these operators, we still have to make them local and 
then pick up the dominant part.  Hence,
\be
{\cal O}^\rho_A(\xi,0)=\sum_n {1 \over n!} \xi^{\mu_1} \cdot \cdot \cdot
\xi^{\mu_n} {\cal O}^\rho_{{\mu_1} \cdot \cdot \cdot {\mu_n}}(0)~.
\ee
As we can see, the above local operator is symmetric in
$\mu_1 ....\mu_n$ but has no definite symmetry in the permutation
of $\rho$ with any of the other indices. The dominant
part of the above operator can be obtained, in the usual twist
analysis, by decomposing into symmetric and mixed symmetric parts, 
\be
{\cal O}^\rho_{\mu_1 \cdot \cdot  \cdot\mu_n}=
{\cal O}^{\{ \rho}_{~~\mu_1  \mu_2\cdot \cdot\cdot  \mu_n\}}
+ {2 n \over n+1} {\cal O}^{\{ [\rho}_{~~\mu_1] \mu_2 \cdot 
\cdot \cdot \mu_n\}}~.
\ee
The fully symmetric part is twist-two 
and the mixed symmetric part
is twist-three. (As usual, $\{~~\}$, $[~~]$ mean symmetrisation
and antisymmetrisation respectively).

Let us now compute the twist-two contribution to $\widetilde W_{\mu \nu}
(p,q,s)$. Expanding the operator matrix element in terms of the 
vectors available in the theory, the most general expression
which is fully symmetric can be written as
\bea
\xi^{\mu_1} \cdot \cdot \cdot \xi^{\mu_n} 
\langle p s| {\cal O}^{\{ \rho}_{~~\mu_1  \mu_2\cdot \cdot\cdot  \mu_n\}}
|p s \rangle
&=&{B_n(p^2) \over (n+1)!} \left[ n! ~ s^\rho (\xi \cdot p)^n 
+n~n! ~ p^\rho \xi \cdot s (\xi \cdot p)^{n-1} \right]~,
\label{MAT1}
\eea
where $B_n(p^2)$ are unknown scalars which contain all the
non-perturbative information.

To perform the integration in eqn.(\ref {WMuNulc}) using the delta function,
we define the function $g(y)$ such that
\be
b(y) ={1\over 2 \pi}\sum_{n=0} {B_n(p^2) \over (n+1)!} 
\int {d(p\cdot \xi) }~ e^{-i y \xi \cdot p} ~(\xi \cdot p)^n ~,
\ee
Substituting eqn.(\ref {MAT1}) in eqn.(\ref {WMuNulc}) and 
using the above Fourier decomposition
we can perform the integrals and the result is
\be
\widetilde W_{\mu \nu}^{(2)}= {i \over 4 ~ {p \cdot q}} 
~ \epsilon_{\mu \nu \lambda \rho} ~ q^\lambda \left[
s^\rho - p^\rho  {s \cdot q  \over  p \cdot q} 
\left(1 + x {d \over dx} \right)  \right] b(x)~.
\ee
Comparing the above expression with eqn.(\ref{WMUNU}), we find
\bea
g_1^{(2)}&=& {1 \over 4} \left ( - x {d \over dx} \right) b(x) ~,\\
g_2^{(2)}&=& {1 \over 4} \left ( 1+x {d \over dx}  \right)b(x) ~,
\eea
where the superscript $2$ denotes the twist of the operators
contributing to the structure functions.
Note that current conservation is maintained, i.e
$q^\mu \widetilde W_{\mu \nu}=0$.
Let us now compute the twist-three contribution to $\widetilde 
W_{\mu \nu}(p,q,s)$.
Again using symmetry arguments we find that
\bea
\xi_{\mu_1} \cdot \cdot \cdot \xi_{\mu_n} 
{\cal O}^{\{ [\rho}_{~~\mu_1] \mu_2 \cdot \cdot \cdot \mu_n\}}
&=& D_n(p^2)~ \xi_{\mu_1} \cdot \cdot \cdot \xi_{\mu_n} ~  
S^{\{[\rho}p^{\mu_1 ]} \cdot \cdot \cdot p^{\mu_n \}} \nonumber \\ 
&=& {D_n(p^2) \over 2} ~ (s^\rho p^\alpha -s^\alpha p^\rho) ~ \xi_\alpha
(\xi \cdot p)^{n-1} ~.
\eea
Substituting this in eqn.(\ref {WMuNulc}) and using
\be
d(y) ={1 \over 2 \pi i} \sum_{n=0} {n  \over (n+1)!} D_n(p^2)
\int {d(p\cdot \xi)}~ e^{-i y \xi \cdot p} ~(\xi \cdot p)^{n-1}~,
\ee
we find that
\bea
\widetilde W_{\mu \nu}^{(3)}&=& {i \over 4 ~ p\cdot q} 
~ \epsilon_{\mu \nu \lambda \rho} ~
\left [q^\lambda \left ({ s.q \over p.q } p^\rho -s^\rho \right ) 
{d \over dx} 
+ p^\lambda s^\rho  \left (1 - x {d \over dx} \right ) \right] d(x) ~,
\label{nomass}
\eea
where the superscript on the hadronic tensor denotes the twist. 
{}From the above equation it is clear that the second term
does not satisfy the current conservation relation. 
It is important to realise that this non-conservation does not
manifest only for the hadronic matrix elements, but continues
to hold even if we were to compute the eqn.(\ref {WMuNulc}) 
between quark states. Doing this, we find
\be
\widetilde W_{\mu \nu}= {i m \over 2 p.q} \epsilon_{\mu \nu \lambda \rho}s^\rho
\left( (p+q)^\lambda \epsilon(q^0+p^0) \delta(1-x)
+(q-p)^\lambda \epsilon(q^0-p^0) \delta(1+x)\right) ~.
\ee
Notice that current conservation is violated even at this
level. It can be maintained if we include the mass term which 
we dropped in the expansion of the time-ordered product.
The mass term turns out to be
\be
S^{(m)}(\xi)=  {1 \over 4 i \pi^2} {m \over \xi^2 - i\epsilon} ~.
\ee
Adding this term to the equation (\ref {prop1}) and  using the equation 
of motion for the quark fields, we get the manifestly current conserved form:
\be
\widetilde W_{\mu \nu}={ 1 \over 8 \pi^2} \epsilon_{\mu \nu \lambda \rho} q^\lambda
\int d^4 \xi e^{iq.\xi} \delta(\xi^2) \epsilon(\xi_0)
\langle p,s \vert : {\cal O}^\rho (\xi,0): \vert p,s\rangle  ~.
\ee
We find that the mass term exactly cancels the current non-conserving 
part appearing in the eqn.(\ref {WMuNulc}) to reproduce the above 
current conserved equation. The above analysis shows us that when we
work at a given order in the twist expansion, it is important to keep
both the singular terms and the regular terms that can contribute at
the given order. The final result that we have when we combine the
twist-2 and the twist-3 contributions is as follows:
\bea
g_1 (x)&=& {1 \over 4 } \left [ - x {d \over dx} ~ b(x)
- {p^2 \over p.q} \left (1+x {d \over dx} \right) d(x)\right] ~,\\
g_2 (x)&=& {1 \over 4 } \left [ \left (1+ x {d \over dx} \right) b(x)
-{d \over dx} ~ d(x) \right] ~.
\eea
Here, one can easily see that $g_1(x,Q^2)$ and $g_2(x,Q^2)$ 
are related by WW sum-rule if the higher twist terms 
such as terms proportional to $p^2$ and $d(x)$ are neglected.      
>From the expression for $g_2(x,Q^2)$, it is interesting to note that the 
twist-two part $b(x)$ and the twist-three part $d(x)$ contribute 
to the cross-section at the same order in $M/\sqrt{Q^2}$. Since
they appear at the same order, it is very difficult to disentangle
these operators and, in general, a measurement of $g_2(x)$ will
be sensitive to both twist-two and twist-three operators at the
same level. While twist-two operators have a simple parton model 
interpretation, higher twist operators cannot be described with
the same simple picture. As a consequence, $g_2(x,Q^2)$ is not
amenable to a parton model interpretation. We shall see, in the
following, that there is yet another problem with the interpretation
of $g_2(x)$, {\it viz.}, the usual factorisation of the hadronic tensor
into hard and soft parts, wherein there is a cancellation of the
infrared and collinear singularities, does not occur for the case
of $g_2$. 
On the contrary, if we consider $g_T$ instead of $g_2$, 
then factorisation can, indeed, be demonstrated. 
In what follows,
we proceed to verify these statements, using the factorisation
method.

\vspace{1.5cm}
  \begin{picture}(1500,9000)
  \drawline\photon[\NW\REG](18000,5000)[8]
  \drawarrow[\E\ATBASE](\pmidx,\pmidy)
  \global\advance\pmidx by 650
  \global\advance\pmidy by 450
  \put(\pmidx,\pmidy){$q$}
  \drawline\fermion[\E\REG](\photonfrontx,\photonfronty)[8000]
  \global\advance\pmidy by 60
  \drawarrow[\E\ATBASE](\pmidx,\pmidy)
  \global\advance\pmidy by 750
  \put(\pmidx,\pmidy){$p^\prime$}
  \drawline\fermion[\SW\REG](\photonfrontx,\photonfronty)[8000]\\
  \drawarrow[\NE\ATBASE](\pmidx,\pmidy)
  \global\advance\pmidx by -550
  \global\advance\pmidy by 450
  \put(\pmidx,\pmidy){$p$}
  \end{picture}

\vspace {.5cm}
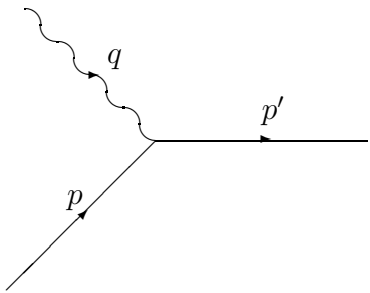
\centerline{\small{Fig.~1. Born diagram.}}
\vspace {.5cm}

	The factorisation theorem \cite{CSS} ensures the separation of long 
distance (soft) effects from the short distance (hard) effects 
and hence in the DIS limit the quark and gluonic contribution to 
the polarised hadron tenor $\widetilde W^{\mu \nu}$ can be 
factorised as

\begin{eqnarray}
\widetilde W^{\gamma^* P}_{\mu \nu} (x,Q^2) &=& \sum_i \int_x^1 
\frac{d y}{y} f_{{\Delta q}/{P}} (y,\mu^2) ~ \widetilde H^{\gamma^* q_i}
_{\mu \nu} (q,y p, \mu^2,\alpha_s (\mu^2))\nonumber \\
&& + \int_x^1 
\frac{d y}{y} f_{{\Delta g}/{P}} (y,\mu^2) ~ \widetilde H^{\gamma^* g}
_{\mu \nu} (q,y p, \mu^2,\alpha_s (\mu^2))~,
\label{FT}
\end{eqnarray}
where $i$ runs over the quark flavours and $\mu$ is the factorisation
scale which defines the separation of short distance from the long 
distance part. 
The soft effects are contained in the parton distribution functions
$f_{\Delta a/P}$ 
which are proton matrix elements of certain gauge invariant bilocal
operators made out of parton fields such as quarks and gluons.
The hard scattering coefficients 
(HSC), $\widetilde H^{\gamma^* a}_{\mu \nu}$ are perturbative and the 
factorisation theorem in the DIS limit ensures that they are free of 
any infrared ($IR$) and collinear singularities and do not depend 
on the properties of the target. This target independence can be
used to advantage: the HSCs can be computed order by order by replacing 
hadron states by asymptotic parton states.  The contribution 
to various structure functions can be extracted by using 
appropriate projection operators.

\vspace{2.0cm}
  \begin{picture}(500,21000)
  \drawline\photon[\NW\REG](10000,18500)[8]
  \drawarrow[\E\ATBASE](\pmidx,\pmidy)
  \global\advance\pmidx by 450
  \global\advance\pmidy by 350
  \put(\pmidx,\pmidy){$q$}
  \drawline\fermion[\E\REG](\photonfrontx,\photonfronty)[8000]
  \global\advance\pmidy by 80
  \drawarrow[\E\ATBASE](\pmidx,\pmidy)
  \global\advance\pmidy by 650
  \put(\pmidx,\pmidy){$p$}
  \drawline\fermion[\S\REG](\photonfrontx,\photonfronty)[8000]
  \drawline\fermion[\E\REG](\fermionbackx,\fermionbacky)[8000]
  \global\advance\pmidy by 80
  \drawarrow[\W\ATBASE](\pmidx,\pmidy)
  \global\advance\pmidy by -1050
  \put(\pmidx,\pmidy){$p^\prime$}
  \drawline\gluon[\SW\FLIPPED](\fermionfrontx,\fermionfronty)[4]
  \global\advance\pmidx by -200
  \global\advance\pmidy by 200
  \drawarrow[\NE\ATBASE](\pmidx,\pmidy)
  \global\advance\pmidx by 950
  \global\advance\pmidy by -950
  \put(\pmidx,\pmidy){$k$}

  \drawline\photon[\NW\REG](30000,18500)[8]
  \drawarrow[\E\ATBASE](\pmidx,\pmidy)
  \global\advance\pmidx by 450
  \global\advance\pmidy by 350
  \put(\pmidx,\pmidy){$q$}
  \drawline\fermion[\E\REG](\photonfrontx,\photonfronty)[8000]
  \global\advance\pmidy by 80
  \drawarrow[\W\ATBASE](\pmidx,\pmidy)
  \global\advance\pmidy by 650
  \put(\pmidx,\pmidy){$p^\prime$}
  \drawline\fermion[\S\REG](\photonfrontx,\photonfronty)[8000]
  \drawline\fermion[\E\REG](\fermionbackx,\fermionbacky)[8000]
  \global\advance\pmidy by 80
  \drawarrow[\E\ATBASE](\pmidx,\pmidy)
  \global\advance\pmidy by -1050
  \put(\pmidx,\pmidy){$p$}
  \drawline\gluon[\SW\FLIPPED](\fermionfrontx,\fermionfronty)[4]
  \global\advance\pmidx by -200
  \global\advance\pmidy by 200
  \drawarrow[\NE\ATBASE](\pmidx,\pmidy)
  \global\advance\pmidx by 950
  \global\advance\pmidy by -950
  \put(\pmidx,\pmidy){$k$}
  \end{picture}

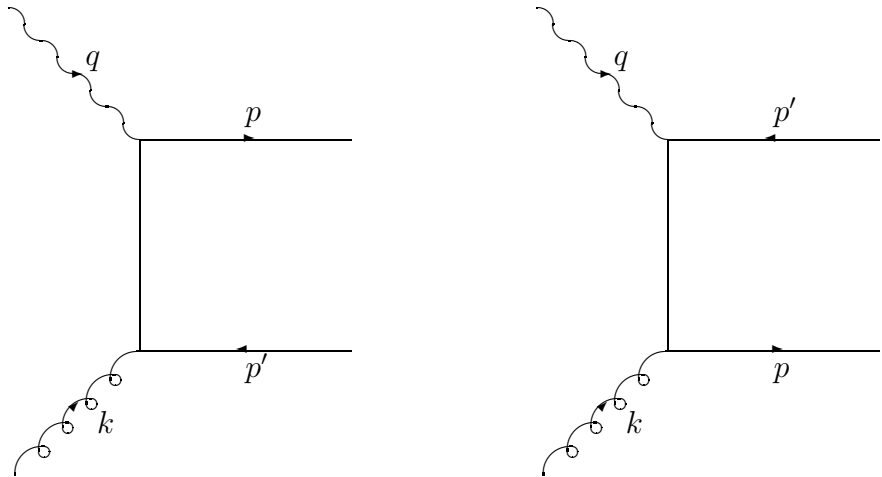
\centerline {\small{Fig.~2. Photon gluon fusion diagram}} 

	Let us begin by evaluating the HSCs to the quark sector to 
leading order.  Replace the proton by quark target ($P \rightarrow q$) 
and retain terms up to order ${\cal O} (\alpha_s^0)$ in eqn.(\ref{FT})
\begin{eqnarray}
\widetilde W^{(0),{\gamma^* q}}_{\mu \nu} (x,Q^2) &=& \sum_i \int_x^1 
\frac{d y}{y} f^{(0)}_{\Delta q/q} (y,\mu^2) ~ \widetilde 
H^{(0),\gamma^* q_i}_{\mu \nu} (q,y p, \mu^2,\alpha_s (\mu^2))~,
\label{FTq}
\end{eqnarray}
where the superscript in the above equation denotes 
the order of strong coupling $\alpha_s$.
$\widetilde W^{(0),{\gamma^* q}}_{\mu \nu}$ to  ${\cal O} (\alpha_s^0)$ 
gets contribution from the Born diagram $\gamma^* q \rightarrow q$ (Fig.~1).
For quarks as target, the distribution function can be calculated from
the operator definitions (see below).  To ${\cal O} (\alpha_s^0)$, $f^{(0)}_{
\Delta q /q} \propto \delta (1-z)$ and hence  $\widetilde H^{(0),\gamma^* q}
_{\mu \nu}$ is same as the Born diagram.  This is the usual statement 
that to leading order the parton model (PM) and the factorisation method 
which is a field theoretical generalisation of the PM are equivalent.

We now discuss factorisation at the next-to-leading order.
To evaluate $\widetilde H^{\gamma^* g}_{\mu \nu}$ in eqn.(\ref{FT}),
we replace $P \rightarrow g$, i.e
\begin{eqnarray}
\widetilde W^{(1),\gamma^* g}_{\mu \nu} (x,Q^2) &=& \sum_i \int_x^1 
\frac{d y}{y} f^{(1)}_{{\Delta q}/{g}} (y,\mu^2) ~ \widetilde 
H^{(0),\gamma^* q_i}_{\mu \nu} (q,y p, \mu^2,\alpha_s (\mu^2))
\nonumber \\
&& + \int_x^1 \frac{d y}{y} f^{(0)}_{{{\Delta g}/{g}}} (y,\mu^2) 
~ \widetilde H^{(1),\gamma^* g}_{\mu \nu} (q,y p, \mu^2,\alpha_s (\mu^2))~,
\label{FTg}
\end{eqnarray}
The LHS is the subprocess cross-section $\gamma^* g \rightarrow q \bar q$
(Fig.~2) and the RHS has two parts $viz.$ the quark and gluon sector.  We 
have shown that $\widetilde H^{(0),\gamma^* q}_{\mu \nu}$ is the subprocess
$\gamma^* q \rightarrow q$. Hence we need to evaluate $f^{(1)}_{\Delta q /g}$
and $f^{(0)}_{\Delta g /g}$ from the definitions given in 
eqn.(\ref{DFq},\ref{DFg}), by replacing $P\rightarrow g$.  To ${\cal O} 
(\alpha_s^0)$, $f^{(0)}_{\Delta g /g} \propto \delta (1-z)$.  To evaluate 
$\widetilde H^{(1),\gamma^* g}_{\mu \nu}$, we have to evaluate 
$\widetilde W^{(1),\gamma^* q}_{\mu \nu}$ and $f^{(1)}_{\Delta q /g}$. 
The above procedure works fine for the unpolarised structure functions 
$F_{1,2} (x,Q^2)$ \cite{CSS} and the longitudinally polarised structure 
function $g_1 (x,Q^2)$ \cite{BQ,AM}, but the transversely polarised structure
function $g_2 (x,Q^2)$ turns out to be an exception.  Projecting the 
contribution to $g_2 (x,Q^2)$ in eqn.(\ref{FTq}), it turns out that $H^{(0),
\gamma^* q}_2 =0$.  This is expected as we know from the PM that $g_2 
(x,Q^2)=0$ to leading order.  As a consequence it turns out from 
eqn.(\ref{FTg}) that $\widetilde W^{(1),\gamma^* g}_2=\widetilde 
H^{(1),\gamma^* g}_2$.  
Hence the usual cancellation of IR and collinear singularities between
the subprocess cross-section and the appropriate parton matrix element, 
to give a HSC free of these singularities does not {\it seem} to occur 
in the case of $g_2 (x,Q^2)$.  The reason for this is that the above 
expression
is incomplete as far as the extraction of $g_2(x,Q^2)$ is concerned.
In fact the operator structure for the $g_2(x,Q^2)$ is much more complicated
than that of $g_1(x,Q^2)$.  Also, the simple minded convolution
may not work for $g_2(x,Q^2)$.  This has earlier been demonstrated 
using free field analysis.  

	Let us now demonstrate the claim of factorisation made for 
the case of $g_T$. To {\it leading twist}, $g_T(x,Q^2)$ gets 
contribution from the parton distributions functions, $viz.$
\begin{eqnarray}
f_{\Delta q /P} (x,\mu^2) &=& \frac{1}{4 \pi} \int d \xi^- e^{-i x \xi^- p^+}
\left [ \langle p s_\perp \vert \bar{\psi}_a (\xi^-) \not\! s_\perp \gamma_5 
{\cal G}^a_b \psi^b (0) \vert p s_\perp \rangle_c \right .\nonumber \\
&& \left . + \langle ps_\perp \vert \bar{\psi}_a (0) \not\! s_\perp \gamma_5 
{\cal G}^a_b \psi^b (\xi^-) \vert p s_\perp \rangle_c \right ]~,
\label{DFq} \\
f_{\Delta g /P} (x,\mu^2) &=& \frac{i}{4 \pi x p^+} \epsilon_{\mu\nu\lambda
\sigma} ~ s^\lambda p^\sigma \int d \xi^- e^{-i x \xi^- p^+} 
\left [\langle ps_\perp \vert F^{+ \mu} (\xi^-) {\cal G}^a_b F^{+ \nu} (0) 
\vert p s_\perp \rangle_c \right . \nonumber \\ 
&& \left . - \langle ps_\perp \vert F^{+ \mu} (0) {\cal G}^a_b F^{+ \nu} 
(\xi^-) \vert p s_\perp \rangle_c  \right ]~,
\label{DFg}
\end{eqnarray}
where the light-cone variables have been used to denote any four vector
$\xi^\mu=(\xi^+,\xi^-,\xi_T)$, with $\xi^\pm=(\xi^0 \pm \xi^3)/\sqrt{2}$.
$F_a^{\mu \nu}$ is the gluon field strength tensor and ${\cal G}_b^a 
\equiv{\cal P} \exp[i g \int_0^{\xi^-} d \zeta^-$ $ A^+ (\zeta^-)]^a_b$ 
is the 
path ordered exponent which restores the gauge invariance of the bilocal 
operators.  We do not consider the other twist-three gluonic operator 
\cite{Jig} that contributes to DIS as they are suppressed by strong 
coupling. For parton targets the above distributions are normalised as
\begin{eqnarray}
f_{(\Delta q + \Delta {\bar q})/a(h)} (z) &=& h ~\delta (1-z) 
~ \delta_{a,(q,\bar q)}~,
\label{NORq} \\
f_{\Delta g /a(h)} (z) &=& h ~\delta (1-z) ~ \delta_{a,g}~,
\label{NORg}
\end{eqnarray}
where $h=\pm 1$ is the helicity of the incoming parton and $z$ is the 
sub-process Bj\"orken variable.

	Using the procedure discussed above, we evaluate the HSCs to
transversely polarised structure function $g_T (x,Q^2)$ and study its 
factorisation properties.  To ${\cal O} (\alpha_s^0)$ we can project 
the $g_T (x,Q^2)$ contribution from eqn.(\ref{FTq}).  The LHS is the 
Born diagram $\gamma^* (q) q(p) \rightarrow q(p^\prime)$ (Fig.~1) and 
its contribution to $g_T (x,Q^2) \neq 0$.  Using the normalisation 
condition eqn.(\ref{NORq}) we find
\begin{eqnarray}
\widetilde H^{(0),\gamma^* q}_T = {e^2  \over 2} ~ \delta (1-z) ~,
\label{HSCq}
\end{eqnarray}
{}From eqn.(\ref{FT}) it is clear that to leading order $g_T (x,Q^2)$ gets 
contribution from the 
parton distribution eqn.(\ref{DFq}).  
At next to leading order $g_T (x,Q^2)$ gets contribution from the other 
parton distribution eqn.(\ref{DFg}).  To evaluate the 
corresponding HSC $\widetilde H^{(1),\gamma^* g}_T (x,Q^2)$, we use 
eqn.(\ref{FTg}) and project out the $g_T (x,Q^2)$ contribution.  This 
involves 
the calculation of the matrix element eqn.(\ref{DFq}) between gluon states 
$f^{(1)}_{\Delta q/g}$ and the sub process cross-section $\widetilde 
W^{(1),\gamma^* g}_T$, both to ${\cal O} (\alpha_s)$.

	The sub-process cross-section $\widetilde W^{(1),\gamma^* g}
_{\mu \nu}$ involves the $\gamma^* (q)g(k) \rightarrow q(p) \bar q
(p^\prime)$ fusion process (Fig 2).  This diagram is free of UV 
divergence but has a mass singularity, which
appears at small scattering angles in the massless limit. We could 
regulate this by keeping either the quark or the gluon mass non-zero,  
but we choose to keep both particles massive as the prescription
dependence would be explicit in this case.  Projecting the contribution 
to $g_T (x,Q^2)$, by using the appropriate projection operator, we get
\begin{eqnarray}
\widetilde W^{(1),\gamma^* g}_T &=& e^2 {\alpha_s \over 4 \pi} 
\int^1_{-1} { d L \over (1-L^2 \omega^2 \kappa)^2} 
\left \{ \left ( -1+z+{k^2 \over q^2} z \right ) (1- L^2 \omega^2)^2
-4 {m^2 \over q^2} z (1+ L^2 \omega^2) \right. \nonumber\\
&+&\!\!\!\! \left. 4{k^2\over q^2} z^2 L^2 \omega^2 \left[ -2(1-z)+4{m^2+k^2 
\over q^2}z -(1-z) (1-L^2 \omega^2) - {k^2\over q^2} z (1+L^2 \omega^2) 
\right]\right\},\nonumber
\end{eqnarray}
where $\omega^2=1-4 m^2/s$, $s=(q+k)^2$ and $L=\cos \theta$, $\theta$ 
being the centre-of-mass scattering angle. Performing the two body 
phase-space integral in the centre-of-mass frame, we get 
\begin{eqnarray}
\widetilde W^{(1),\gamma^* g}_T =e^2 ~ {\alpha_s \over 2 \pi} ~ \frac
{k^2 z (1-z)^2} {m^2-k^2 z (1-z)} ~,
\label{Wg}
\end{eqnarray}
Note that the contribution to $g_T (x,Q^2)$ is independent of the $\ln 
Q^2$ term.  Both $g_1 (x,Q^2)$ and $g_2 (x,Q^2)$ separately 
depend on the $\ln Q^2$, but the combination $g_T (x,Q^2)$ is 
independent of the $\ln Q^2$ term.
The constant piece depends on the choice of regulator
and hence is prescription dependent as is clearly seen.

\vspace{1cm}

  \begin{picture}(2000,16000)
  \drawline\fermion[\E\REG](5000,12750)[6000]
  \drawline\fermion[\E\REG](5000,13000)[3000]
  \drawline\fermion[\N\REG](\fermionbackx,\fermionbacky)[200]
  \drawline\fermion[\S\REG](\fermionfrontx,\fermionfronty)[400]
  \drawline\fermion[\E\REG](\fermionfrontx,\fermionfronty)[3000]
  \drawline\fermion[\NW\REG](\fermionbackx,\fermionbacky)[4000]
  \global\advance\fermionbackx by -400
  \global\advance\fermionbacky by -100
  \global\advance\pbackx by -1250
  \global\advance\pbacky by +1550
  \put(\pbackx,\pbacky){$\gamma_\perp \gamma_5$}
  \put(\fermionbackx,\fermionbacky){$\bullet$}
  \global\advance\pmidx by 450
  \global\advance\pmidy by 350
  \put(\pmidx,\pmidy){$r$}
  \drawline\fermion[\S\REG](\fermionfrontx,\fermionfronty)[6000]
  \drawarrow[\N\ATBASE](\pmidx,\pmidy)
  \drawline\fermion[\W\REG](\fermionbackx,\fermionbacky)[3000]
  \drawline\gluon[\SE\REG](\fermionfrontx,\fermionfronty)[4]
  \global\advance\pmidx by 400
  \global\advance\pmidy by 150
  \drawarrow[\NW\ATBASE](\pmidx,\pmidy)
  \global\advance\pmidx by 650
  \put(\pmidx,\pmidy){$k$}
  \drawline\fermion[\N\REG](\fermionbackx,\fermionbacky)[200]
  \global\advance\pmidy by 550
  \put(\pmidx,\pmidy){$p$}
  \drawline\fermion[\S\REG](\fermionfrontx,\fermionfronty)[200]
  \drawline\fermion[\W\REG](\fermionfrontx,\fermionfronty)[3000]
  \drawline\fermion[\N\REG](\fermionbackx,\fermionbacky)[6000]
  \drawarrow[\S\ATBASE](\pmidx,\pmidy)
  \drawline\gluon[\SW\FLIPPED](\fermionfrontx,\fermionfronty)[4]
  \global\advance\pmidx by -500
  \drawarrow[\NE\ATBASE](\pmidx,\pmidy)
  \global\advance\pmidx by -750
  \put(\pmidx,\pmidy){$k$}
  \drawline\fermion[\NE\REG](\fermionbackx,\fermionbacky)[4300]
  \global\advance\pmidx by -650
  \global\advance\pmidy by 450
  \put(\pmidx,\pmidy){$r$}
  \drawline\fermion[\N\REG](\fermionbackx,\fermionbacky)[400]
  \drawline\fermion[\S\REG](\fermionfrontx,\fermionfronty)[600]

  \drawline\fermion[\E\REG](29000,12750)[6000]
  \drawline\fermion[\E\REG](29000,13000)[3000]
  \drawline\fermion[\N\REG](\fermionbackx,\fermionbacky)[200]
  \drawline\fermion[\S\REG](\fermionfrontx,\fermionfronty)[400]
  \drawline\fermion[\E\REG](\fermionfrontx,\fermionfronty)[3000]
  \drawline\fermion[\NW\REG](\fermionbackx,\fermionbacky)[4000]
  \global\advance\fermionbackx by -400
  \global\advance\fermionbacky by -100
  \global\advance\pbackx by -1250
  \global\advance\pbacky by +1550
  \put(\pbackx,\pbacky){$\gamma_\perp \gamma_5$}
  \put(\fermionbackx,\fermionbacky){$\bullet$}
  \global\advance\pmidx by 450
  \global\advance\pmidy by 350
  \put(\pmidx,\pmidy){$r$}
  \drawline\fermion[\S\REG](\fermionfrontx,\fermionfronty)[6000]
  \drawarrow[\S\ATBASE](\pmidx,\pmidy)
  \drawline\fermion[\W\REG](\fermionbackx,\fermionbacky)[3000]
  \drawline\gluon[\SE\REG](\fermionfrontx,\fermionfronty)[4]
  \global\advance\pmidx by 400
  \global\advance\pmidy by 150
  \drawarrow[\NW\ATBASE](\pmidx,\pmidy)
  \global\advance\pmidx by 650
  \put(\pmidx,\pmidy){$k$}
  \drawline\fermion[\N\REG](\fermionbackx,\fermionbacky)[200]
  \global\advance\pmidy by 550
  \put(\pmidx,\pmidy){$p$}
  \drawline\fermion[\S\REG](\fermionfrontx,\fermionfronty)[200]
  \drawline\fermion[\W\REG](\fermionfrontx,\fermionfronty)[3000]
  \drawline\fermion[\N\REG](\fermionbackx,\fermionbacky)[6000]
  \drawarrow[\N\ATBASE](\pmidx,\pmidy)
  \drawline\gluon[\SW\FLIPPED](\fermionfrontx,\fermionfronty)[4]
  \global\advance\pmidx by -500
  \drawarrow[\NE\ATBASE](\pmidx,\pmidy)
  \global\advance\pmidx by -750
  \put(\pmidx,\pmidy){$k$}
  \drawline\fermion[\NE\REG](\fermionbackx,\fermionbacky)[4300]
  \global\advance\pmidx by -650
  \global\advance\pmidy by 450
  \put(\pmidx,\pmidy){$r$}
  \drawline\fermion[\N\REG](\fermionbackx,\fermionbacky)[400]
  \drawline\fermion[\S\REG](\fermionfrontx,\fermionfronty)[600]
  \end{picture}

\vspace{.5cm}
\centerline {\small Fig.~3. 
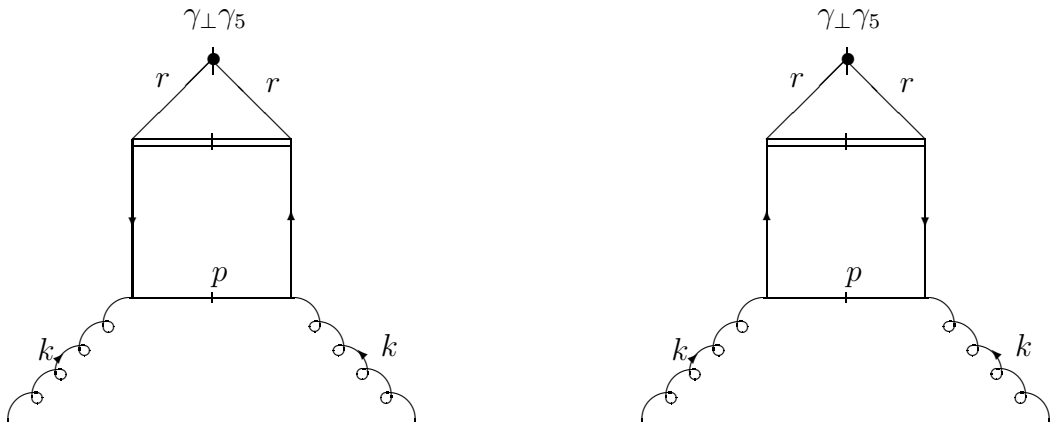
The ${\cal O} (\alpha_s)$ contribution to the matrix $f_{\Delta q/g}$.} 
\vspace{.5cm}

	The matrix element $f^{(1)}_{\Delta q/g}$ (Fig.~3) is evaluated 
using the parton distribution eqn.(\ref{DFq}) in the light-cone gauge 
with the replacement
$P \rightarrow g$.  We keep the quarks and gluon off mass-shell as we did
for the evaluation of cross-section.  Noting that this matrix element
is superficially divergent, we compute it using dimensional regularisation
method, and the matrix element turns out to be
\be
f^{(1)}_{(\Delta q+\Delta \bar q)/g }= 2 \alpha_s \int {d^{d-2} p_{\perp} 
\over (2 \pi)^{d-2}} ~ {1-z \over (p_\perp^2+m^2 - k^2 z (1-z))^2 } 
\left [ m^2 + k^2 z (1-z) + {d-4 \over d-2} p_\perp^2 \right] ~.
\ee   
As is clear, the integral is convergent and reduces to a simple form:
\begin{eqnarray}
f^{(1)}_{(\Delta q+ \Delta \bar q)/g} = {\alpha_s \over \pi} ~ \frac
{k^2 z (1-z)^2} {m^2-k^2 z (1-z)} ~.
\label{PDg1}
\end{eqnarray} 
The term $(d-4)/(d-2)$ in the integral gives non-vanishing
finite contribution.  We also checked the correctness of our result in the
Pauli-Villars (PV) regularisation scheme.  We could reproduce
the same result in this scheme also confirming that our 
finite result is UV scheme independent.  In the PV regularisation,
since the integral is performed in four dimension, the
$(d-4)/(d-2)$ term is absent.  The analogous term comes from  
the integral with $m$ replaced by $M$ (PV regulator) in the 
limit $M$ goes to infinity.  
Hence, our result is independent of UV scheme.   
This is not true in the case
of operator matrix elements which one encounters in the evaluation
of the structure functions $F_2(x,Q^2)$ and $g_1(x,Q^2)$ \cite {BQ}.
Recall that the matrix elements appearing in the evaluation of the 
QCD corrections to $F_2(x,Q^2)$ and $g_1(x,Q^2)$ are UV renormalisation 
scheme dependent.  In other words, those operators 
are defined/renormalised in a definite UV renormalisation 
scheme say $\overline {MS}$ or momentum subtraction scheme or 
Pauli-Villars scheme.  
In our case, since the matrix element is finite the result is
UV renormalisation scheme independent to this order.
Observe that the masses we introduced to avoid IR singularities
lead to two different results when one considers the cross-section
and the matrix element separately.   That is, both the cross-section
and the matrix element are dependent on the order in which 
the masses go to zero.  This is the usual prescription dependence
one encounters in massless theories. 
The prescription-dependent structure of the above equation is the same as 
that of $W^{(1),
\gamma^*g}$.  Substituting for the normalisation condition eqn.(\ref
{NORg}) and eqn.(\ref{HSCq},\ref{Wg},\ref{PDg1}) in eqn.(\ref{FTg}), we get
\begin{eqnarray}
\widetilde H^{(1),\gamma^* g}_T = 0 ~.
\label{HSCg}
\end{eqnarray}
Note that there is a cancellation of the prescription dependent pieces
confirming the factorisation.  Also, it turns out that the next to 
leading order HSC, $\widetilde H^{(1),
\gamma^* g}_T$ is zero and hence twist-three distribution eqn.(\ref{DFg})
does not contribute to $g_T (x,Q^2)$ to any of the moment to this order.  
The above analysis proves that the first moment of the gluon coefficient 
function is zero in FM.

One of the important outcomes of the demonstration of factorisation
for $g_T$ is that it admits a description in terms of a process-independent
universal distribution. This distribution is no longer a parton distribution
in the usual sense of the term, because of the twist-three contribution
to $g_T$. But the process-independence is still useful, so that once the
non-perturbative distribution associated with $g_T$ has been extracted
in one experiment (in DIS, for example), it can be used to make predictions
for other processes, like Drell-Yan. 

The above analysis also has interesting consequences for the first moment
of the structure function $g_T(x,Q^2)$.  At the leading order, one 
would be led by the validity of the BC sum-rule to conclude that
the first moment of $g_T(x,Q^2)$ is same as that of $g_1(x,Q^2)$.  
Though these first moments are measured in completely different experiments
($g_T(x,Q^2)$ in transversely polarised DIS and $g_1(x,Q^2)$ in longitudinally
polarised DIS), they should coincide numerically. Note that because 
the BC sum-rule is valid at one loop order \cite{BC1}, one would
expect that the first moment of the gluonic coefficient vanishes in the FM.
Our analysis of $g_T$ using the factorisation method confirms this
expectation. The first moments of both $g_1(x,Q^2)$ and $g_T(x,Q^2)$ 
in FM, are related to one and the same matrix element $\langle p s|\bar 
\psi \not \! s \gamma_5 \psi |p s \rangle$ which is Lorentz invariant.  
The argument used here is the same as the rotational invariance argument 
that may be used to justify the BC sum-rule \cite{Fey,Jaf-R}.  

A related issue is that of the hard gluonic contribution to the
first moment of $g_1$ $via$ the anomaly \cite{anomaly}. This gluonic
contribution induced through the anomaly is, in fact, a possible
explanation for the surprisingly small value for the first moment
of $g_1$ measured in experiments \cite{g1}. In the FM \cite{BQ}, 
however, this contribution vanishes as long as the quark distributions 
are related to matrix elements of the standard quark field-operators
which appear in the operator product expansion \cite{JM}. In a parton model
computation of the hard gluonic contribution due to the anomaly,
the gluonic coefficient is found to be non-zero \cite{anomaly}.
In other words, the size of the gluonic contribution is dependent on 
the definition of the parton distributions. In the case where there is 
a non-vanishing hard gluonic contribution to the first moment of $g_1$ 
induced by the anomaly, our analysis would tell us that precisely the 
same contribution will also affect $g_T$. Thus, a measurement of the
first moment of $g_T (x)$ will provide a very interesting cross-check
about the importance of the anomaly-induced gluonic contribution.

Another interesting prediction for $g_T$ could be the analogue of the
Bj\" orken sum rule for $g_1$. Given that the first moments of $g_1$
and $g_T$ are identical, we would expect that $g_T$ would satisfy a
sum-rule which is exactly the same as the Bj\" orken sum-rule, and
whose numerical value is the same as that for $g_1$.

In conclusion, we have shown that the transverse structure function 
$g_T(x,Q^2)$($g_1(x,Q^2)+g_2(x,Q^2)$) contains a simple operator structure
which renders one to understand the spin structure of the proton
from a completely different experiment involving transversely polarised proton.  
In addition, due to the simplicity in its structure, 
it is easier to extract and hence understand the higher twist effects.
We have shown that the first moment of $g_T(x,Q^2)$ measures
the spin contributions coming from various partons to the proton spin
using the Factorisation method.  The interesting
point to observe is that at large $Q^2$ to order $\alpha_s(Q^2)$,
with appropriate operator definitions for transverse partons
inside the transversely polarised proton, the factorisation of
mass singularities works.  We have found that the gluonic contribution
to $g_T(x,Q^2)$ is zero to this order for the operators discussed.

\vspace{.7cm}
\noindent
{\Large \bf Acknowledgements:}

PM and VR would like to thank G. Ridolfi and Diptiman Sen for clarifying some 
of the techniques used in the paper.  KS would like to thank R.G. Roberts for 
useful discussions.  VR thanks M.V.N. Murthy for his constant encouragement.  
We thank the organisers of WHEPP-IV for warm hospitality where 
part of this work was done.

\eject

\end{document}